\documentclass[12pt]{article}
\author{A.V.\,Samsonov}
\title{Decay constant of the $\eta$-meson \\ from QCD sum rule}

\begin{document}
\date{}
\maketitle
\it
$$\centerline{\hbox{Institute of Theoretical and Experimental
Physics}}$$
$$\centerline{\hbox{B. Cheremushkinskaya 25, 117259, Moscow, Russia}}$$
  \\ \\

\bf
$$\centerline{\hbox{Abstract}}$$
\rm
\indent Decay constant of the $\eta$-meson $f_\eta$ is calculated on the 
basis of the QCD sum rule method. An instanton contribution is taken into
account. The result is $f_\eta=0.17\pm 0.01\,GeV$.
\\ \\ \\
\large{
\indent The quantum chromodynamics (QCD) sum rule technique was originally
suggested by M.\,Shifman, A.\,Vainshtein and V.\,Zakharov [1]. According to 
this technique the operator product expansion of the pola\-rization operator 
for various currents is considered. The terms of this expansion are the 
vacuum expectation values of different operators, such as quark condensate 
$\langle0\vert\overline qq\vert0\rangle$, gluon condensate 
$\langle0\vert G_{\mu \nu}^a G_{\mu \nu}^a\vert0\rangle$ and other operators 
with higher dimensions, multiplied by calculated in QCD factors.
Such way the non-perturbative corrections in QCD 
are taken into account. Perturbative contribution is also included.
On the other hand, polarization operator is expressed 
phenomenologically via the characteristics of physical states. In order to 
obtain sum rule, both representations are equated to each other after the
Borel transformation $B_{M^2}$:
$$B_{M^2}f(Q^2)=\lim_{n,\,Q^2\to\infty,\;Q^2/n=M^2} {(Q^2)^{n+1}\over n!}
\biggl(-{d\over{dQ^2}}\biggr)^n f(Q^2)\;,\eqno(1)$$
where $M^2$ is the Borel parameter.\\
\indent Decay constant of the $\eta$-meson $f_\eta$ is defined in the 
following way:
$$\langle0\vert j_{\mu 5}\vert\eta(p)\rangle=if_\eta p_\mu\;,$$
where $j_{\mu 5}$ is the axial vector current with $\eta$-meson quantum 
numbers:
$$j_{\mu 5}={1\over\sqrt6}(\overline u\gamma_\mu \gamma_5u+\overline d
\gamma_\mu \gamma_5d-2\overline s\gamma_\mu \gamma_5s)\;.\eqno(2)$$
\indent In order to calculate $f_\eta$ on the basis of QCD sum rule we 
consider pola\-rization operator $\Pi_{\mu\nu}(q)$:
$$\Pi_{\mu\nu}(q)=i\int{d^4x\,e^{iqx}
\langle0\vert T(j_{\mu 5}(x)\,j_{\nu 5}(0))\vert 0\rangle}\;,\eqno(3)$$
where $j_{\mu 5}$ is defined in (2).
Polarization operator has the following structure:
$$\Pi_{\mu\nu}(q)=g_{\mu\nu}\Pi_L(q^2)+(q_\mu q_\nu-q^2g_{\mu\nu})\Pi_T(q^2)
\;.$$
Then:
$$\Pi_{\mu\nu}(q)q_\mu q_\nu=q^2 \Pi_L(q^2)\;.$$
\indent We assume that the masses of $u,\,d$ quarks are zero, but the mass 
of $s$ quark has the non-zero value, $m_s(1GeV)=0.15\,GeV$. \\
\indent We consider negative $q^2$:
$q^2<0\;,\;\;-q^2>>R_c^{-2}$, where $R_c$ is the confinement radius. \\
\indent For polarization operator (3) with current (2) one can obtain:
$$\Pi_{\mu\nu}q_\mu q_\nu =
{16\over3}m_s\langle0\vert\overline ss\vert 0\rangle+i\int{d^4x\,e^{iqx}
\langle0\vert T(\partial j_5(x)\,\partial j_5(0))\vert 0\rangle}\;.$$
In this formula $\partial j_5=-4/\sqrt6\,im_s \overline s\gamma_5s$ 
is the pseudoscalar current.
The polarization operator for the pseudoscalar current was obtained in [1].
For our choice of $j_{\mu 5}$ we have:
$$\displaylines{\Pi_L=-{16\over3}
{{m_s\langle0\vert\overline ss\vert 0\rangle}\over{Q^2}}+
{m_s^2\over{\pi^2}}\ln{Q^2\over{\mu^2}}+\hfill\cr}$$
$$\displaylines{\hfill
+{8\over3}m_s^2
\biggl[-{{m_s\langle0\vert\overline ss\vert 0\rangle}\over Q^4}+
{{\alpha_s\langle0\vert G_{\mu \nu}^a G_{\mu \nu}^a\vert0\rangle}\over
{8\pi Q^4}}+
{112\over 27}{{\pi \alpha_s{\langle0\vert\overline ss\vert 0\rangle}^2}\over
{Q^6}}+I(Q^2)\biggr]\;,\;\;(4)\cr}$$
where $Q^2=-q^2$, $\mu$ is the normalization scale and $I(Q^2)$ is the 
instanton term. In the pseudoscalar channel the direct instanton contribution
is not small. It was shown by B.V.\,Geshkenbein and B.L.\,Ioffe [2], also see
[3,4].\\
\indent Now we represent polarization operator phenomenologically as a sum
of $\eta$-meson and continuum contributions:
$$\Pi_L(q^2)={f_\eta^2q^2\over{m_\eta^2-q^2}}+{1\over\pi}
\int\limits_{s_0}^\infty{{c(t)dt\over{t-q^2}}}\;.\eqno(5)$$
In this formula $m_\eta$ is
the mass of $\eta$-meson, $s_0$ is the continuum threshold. 
Such representation is typical for polarization operator in the sum rule 
method. It incorporates its main features: 
existence of some lowest resonance and approach to the perturbative 
result at large $Q^2$, as demanded by asymptotic freedom. \\
\indent After the substitution (5) into equation (4) and Borel transformation
(1) we obtain:
$$\displaylines{
f_\eta={\sqrt8\over{\sqrt3 m_\eta}}\exp({m_\eta^2\over{2M^2}})
\times\hfill\cr}$$
$$\displaylines{\hfill\times
\biggl[2a+m_s^2(\mu^2)L^{-{8\over9}}\biggl({b_1\over{M^2}}
+{b_2\over{M^4}}-{3\over{8\pi^2}}M^2(1-\exp(-{s_0\over{M^2}}))
+\tilde I(M^2)\biggr)\biggr]^{1/2}\;,\;\;\;(6)\cr}$$
where 
$$ a=-m_s\langle0\vert\overline ss\vert 0\rangle,\;\;
b_1=a+{{\alpha_s\langle0\vert G_{\mu \nu}^a G_{\mu \nu}^a\vert0\rangle}\over
{8\pi}}\,,\;\;
b_2={56\over{27}}\pi\alpha_s{\langle0\vert\overline ss\vert 0\rangle}^2\,,$$
$$L=\ln{M\over\Lambda}/\ln{\mu\over\Lambda}\,.$$
$\tilde I(M^2)$ is the instanton contribution. It calculated in the single 
instanton approximation [4,5,6].
$$\tilde I(M^2)={3\sqrt2\over{8\pi^2}}\xi_\eta\rho_c(M^2)^{3/2}
\int\limits_0^\infty{\int\limits_0^\infty{dxdy\,w
\exp(-{1\over4}\rho_c^2w^2M^2)}}\,,$$
where $$w=\sqrt{1+x^2}+\sqrt{1+y^2}\;\;\hbox{and}\;\; \xi_\eta=
1-{4\over3}\;{1\over{1+{{2\pi\rho_c\sqrt{n_c}}\over{\sqrt3m_s}}}}\,.$$ 
Instanton density has the form [4]: 
$$dn/d\rho=n_c\,\delta(\rho-\rho_c)\,,$$ 
where $\rho_c$ and $n_c$ are two parameters, instanton radius and total 
instanton density. Instanton radius $\rho_c\approx 0.3\,fm$ is assumed 
relatively small compared average instanton separation, which is 
approximately equal to $1\,fm$.\\  
\indent The parameters in (6) have the following values: \\
$\mu=1\,GeV,$\\
$\Lambda=0.2\,GeV,$\\
$m_\eta=0.55\,GeV,$\\
$m_s(1\,GeV)=0.15\,GeV,$\\
$\langle0\vert\overline ss\vert0\rangle=
0.7\langle0\vert\overline qq\vert0\rangle=0.97\cdot10^{-2}\,GeV^3$
(This value is considered in [7,4]),\\
${\alpha_s\over\pi}\langle0\vert G_{\mu \nu}^a G_{\mu \nu}^a\vert0\rangle=
0.012\,GeV^4,\\$
$s_0=2.5\,GeV^2,$\\
$\rho_c=1.67\,GeV^{-1},$\\
$n_c=0.001\,GeV^4.$ \\
The values of $\rho_c$ and $n_c$ are discussed in [6].\\
\indent Substituting these values into equation (6), we obtain $f_\eta$ as
 the function of $M^2$. This function is shown in fig.\,1. The range of $M^2$
is limited by two conditions: the continuum contribution does not exceed
 40-50\% and the power corrections don't exceed 10-15\%. In our case these
conditions give too wide interval of $M^2$. We use the typical values:
$0.9\,GeV^2<M^2<1.3\,GeV^2$. In this interval we obtain:
$$f_\eta=0.17\pm 0.01\,GeV.$$
\indent This result is in a good agreement with the values of $f_\eta$ 
from number of papers [8,9,10].\\
\indent It is worth noting, however, that Particle Data Group [11] gives the
different value of $f_\eta$: $f_\eta^{PDG}=0.13\pm 0.01\,GeV.$
\\ \\ \\
\indent I am grateful to prof.\,B.L.\,Ioffe for stating the problem and 
helpful discussions.
\\ 
\newpage
$\;\;$\\
\centerline{\Large{References.}}\\ \\
$[1]$ M.\,Shifman, A.\,Vainshtein and V.\,Zakharov, Nucl.Phys. B147 (1979)\\
\indent 385, 448.\\
$[2]$ B.V.\,Geshkenbein and B.L.\,Ioffe, Nucl.Phys. B166 (1980) 340.\\
$[3]$ V.\,Novikov, M.\,Shifman, A.\,Vainshtein and V.\,Zakharov, Nucl.Phys.\\
\indent B191 (1981) 301.\\
$[4]$ E.\,Shuryak, Nucl.Phys. B214 (1983) 237.\\
$[5]$ E.\,Shuryak, Nucl.Phys. B203 (1982) 93, 116.\\
$[6]$ T.\,Schafer, E.\,Shuryak, hep-ph/9610451.\\
$[7]$ B.L.\,Ioffe, Nucl.Phys. B188 (1981) 317.\\
$[8]$ Th.\,Feldmann, P.\,Kroll and B.\,Stech, hep-ph/9802409.\\
$[9]$ H.\,Leutwyler, hep-ph/9709408.\\
$[10]$ B.L.\,Ioffe, A.G.\,Oganesyan, Phys.Rev. D57(1998) R6590.\\
$[11]$ R.\,Barnett et al, Particle Data Group, Phys.Rev. D54 (1996) 1.
\newpage
\unitlength=1.00mm
\special{em:linewidth 0.4pt}
\linethickness{0.4pt}
\begin{picture}(120.00,130.00)
\put(50.00,80.00){\vector(1,0){70.00}}
\put(118.00,75.00){\makebox(0,0)[cc]{\small{$M^2,GeV^2$}}}
\bezier{176}(60.00,112.00)(76.00,100.00)(100.00,97.00)
\put(50.00,80.00){\vector(0,1){50.00}}
\put(60.00,78.00){\line(0,1){4.00}}
\put(70.00,82.00){\line(0,-1){4.00}}
\put(80.00,78.00){\line(0,1){4.00}}
\put(90.00,82.00){\line(0,-1){4.00}}
\put(100.00,78.00){\line(0,1){4.00}}
\put(48.00,90.00){\line(1,0){4.00}}
\put(52.00,100.00){\line(-1,0){4.00}}
\put(52.00,120.00){\line(-1,0){4.00}}
\put(48.00,110.00){\line(1,0){4.00}}
\put(70.00,75.00){\makebox(0,0)[cc]{\small{1.0}}}
\put(80.00,75.00){\makebox(0,0)[cc]{\small{1.1}}}
\put(90.00,75.00){\makebox(0,0)[cc]{\small{1.2}}}
\put(100.00,75.00){\makebox(0,0)[cc]{\small{1.3}}}
\put(60.00,75.00){\makebox(0,0)[cc]{\small{0.9}}}
\put(43.00,90.00){\makebox(0,0)[cc]{\small{0.15}}}
\put(43.00,100.00){\makebox(0,0)[cc]{\small{0.16}}}
\put(43.00,110.00){\makebox(0,0)[cc]{\small{0.17}}}
\put(43.00,120.00){\makebox(0,0)[cc]{\small{0.18}}}
\put(41.00,129.00){\makebox(0,0)[cc]{\small{$f_\eta,GeV$}}}
\put(80.00,65,00){\makebox(0,0)[cc]{Fig.\,1.
 Function $f_\eta=f_\eta(M^2)$.}}
\end{picture}
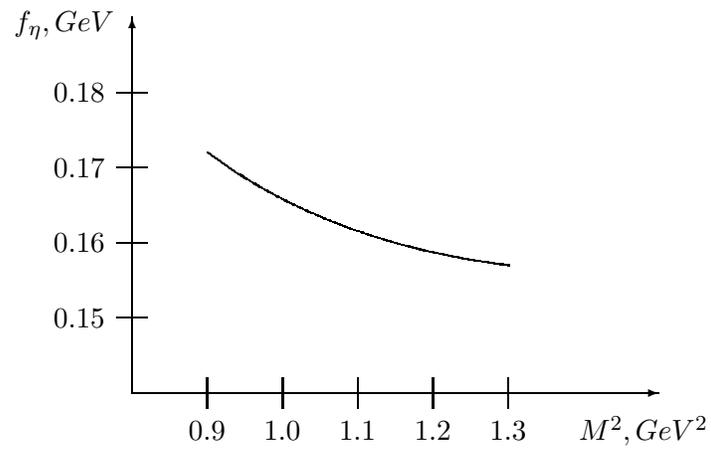
\end{document}